\documentclass[11pt]{article}
\usepackage{float}
\usepackage{array}
\usepackage{cool}
\usepackage{bbold}
\usepackage{eulervm}
\usepackage{hyperref}
\usepackage{tabularx}
\hypersetup{colorlinks=true,linkcolor=blue,citecolor=blue}
\let\oldmb\mathbold
\protected\def\mathbold{\oldmb}
\usepackage{amsmath}
\usepackage{todonotes}
\usepackage{graphicx}
\usepackage{subfig}
\usepackage{color}
 \usepackage{multicol}
\begin{document}
\title{\bf Faraday effect optical sensing of single-molecules by graphene-based layered structures}

\author{D. Jahani \footnote{\href{d.jahani@sharif.edu}{d.jahani@sharif.edu}}, O. Akhavan, A. Alidoust Ghatar,  W. Fritzsche and F. Garwe}
\maketitle {\it \centerline{
\emph{Department of Physics, Sharif University of Technology, P.O. Box 11155-9161, Tehran, Iran}}
\maketitle {\it \centerline{
\emph{ Leibniz Institute of Photonic Technology (IPHT), Jena, Germany. }}

\begin{abstract}
\emph{Recently, modulation of the energy bandgap of graphene when gas molecules are adsorbed to its surface has been proved to be possible. Motivated by this, based on numerical calculations, we investigate the effect of the associated bandgap opening in graphene's spectrum on the sensing properties of the Faraday rotation (FR) of linearly polarized electromagnetic modes in a 1D photonic graphene-based sensor with a microcavity defect channel covered by two graphene layers. Our proposed model introduces an optical contact-free mechanism for the detection of low-concentration molecules attached to graphene's surface. We also show that FR angles could reveal reversal signs for a different amount of surface transfer doping. }
\end{abstract}
\vspace{0.5cm} {\it \emph{Keywords}}: \emph{Single-molecule detection; Faraday rotation; Graphene; Photonic Crystal; Fluidic microchannel.}
%\begin{multicols}{2}
\section{Introduction}
 \emph{Graphene, a single sheet of carbon atoms arranged in a honeycomb structure, was isolated experimentally for the first time from graphite in 2004 [1,2,3]. Thanks to its massless Dirac charge carries, this one-atom-thick material mimics the properties of a gapless semiconductor and has shown outstanding electronic and optical properties even at room temperature [4,5,6]. These interesting properties make graphene serve as an ideal candidate for real-world technologies. Recently, graphene has also shown huge attention-getting achievements as an additional key element in many sensing applications [7,8,9,10]. Surprisingly, because of the low-noise nature of massless charge carriers of graphene, it an ideal candidate for the detection of individual gas molecules adsorbed on its surface [11]. Then, the exceptional electronic property allows the detection of a few molecules on graphene upon surface transfer doping which correspondingly changes the electrical conductivity of graphene in order to be measured by proper electrical contacts. In addition to its unique electrical properties, graphene also reveals peculiar optical characteristics including, high transparency and broadband ultra-fast response to the electromagnetic radiation which could interact with its massless Dirac fermions [12,13]. Consequently, graphene is used in optical contact-free sensing methods for a broad range of light frequencies [14,15,16,17]. Therefore, optical spectroscopies in the presence of graphene provide a highly-sensitive nonintrusive method toward the chip-scale sensing technology in different branches of science. In this regard, in biotechnology, photonic graphene-based biosensors operating in a relatively low-frequency background of light prorogation have received considerable attention in recent years [18,19]. Interestingly, the optical sensing properties of graphene could be enhanced by applying a constant magnetic field normal to its surface. In fact, due to induced optical excitations in the surface conductivity of graphene, applying a magnetic field could rotate the polarization plane of the impinging light [20]. This finite rotation of the polarization's plane of the light propagating through a dielectric medium under Hall situation is called Faraday rotation (FR) which could be also applied for remote sensing of magnetic field intensities. Surprisingly, large FR under the Hall regime has been reported for graphene in spite of being just a one-atom-thick atomic layer [21,22]. This FR in graphene occurs due to intra- and inter-band excitations in its surface optical conductivity which could be controlled by the magnetic field intensity [23]. However, there is a trade-off between FR and the transmission's height of the transmitted light and, therefore, the possibility of finding simultaneously both large FR angles and relatively high transmissions is of high interest in graphene-based integrated optical devices [24,25,26].}
\par
\emph{Integrated optical technology has provided a broad range of opportunities for graphene to be used in sensing applications [27,28]. In fact, the presence of graphene in these devices could improve the sensitivity of optical contact-free methods which is of prime interest in biotechnology [29,30]. Moreover, tunability is another important feature for detection purposes since it suggests more control over the sensing process in optical instruments. In this regard, 1D photonic crystals (PCs), periodically multilayer structures that are designed to control the light prorogation could give more tunability features due to the forbidden frequency zone in the corresponding light spectrum [31]. Interestingly, creating a defect in these optical structures could induce a defect mode within the forbidden region which opens a more controllable route for electromagnetic waves prorogation. Fortunately, growing successes in photonics applications provide cost-effective components for PCs in order to host graphene in chemical and biosensing devices. Now, the effect of the magnetic field on the plane of the polarization of EM waves in a 1D PC could serve for the detection of gas molecules as dopants on the graphene's surface which might show a Dirac gap in its linear dispersion relation. To the best of our knowledge still little has been reported on FR in defective PCs for sensing applications in spite of the fact that the Faraday effect itself is employed for remote sensing of magnetic fields. Moreover, the surface conductivity of graphene in a magnetic field upon absorption which could also lead to the creation of Dirac gaps gives different responses to decoupled polarizations and, therefore, the angle of the rotation could be affected by the absorption of molecules [32].}
\par
\emph{The first incorporation of graphene into a 1D photonic crystal was introduced by O. Berman [33,34]. Since then graphene has been offered to be employed in 1D PCs in order to enhance the optical properties of these layer structures. For instance, the effect of a magnetic field on the polarization plane of defect states emerged into a 1D graphene-based defective PC proved to enhance the Faraday effect for which the plane of the polarization of light is rotated while passing into a magnetic material [35]. Interestingly, the Faraday effect in graphene-based PCs could be considered in optical sensing devices. In this work, we address the magneto-sensing properties of a 1D optical structure composed of a fluidic channel covered by two-graphene layers. In fact, it is the presence of graphene-coated on internal walls of the fluidic channel which leads to the rotation of the linearly polarized light considered to be an equal combination of left- and right-handed circular modes. Note that, to be applicable in real-world applications, the Faraday effect must be accompanied by decently high transmission peaks [36]. Also, the realization of a Dirac gap upon coating graphene and also applying a strong magnetic field normal to its surface is expected. Therefore, the effect of opening Dirac gaps in graphene's spectrum considering the low Fermi energy's chances which occur due to the absorbed gas molecules on the graphene's surface on FR angles is worth huge attention. To the best of our knowledge, to this stage, no study has been reported  on the effect of opening a Dirac gap on the Faraday effect of circular modes in a 1D graphene-based defective PC.}
\par
\emph{This paper is organized into the following sections: The model and theory is introduced in section 2. Numerical calculations are addressed in section 3 where we illustrate the effect of the opening an electronic bandgap on the low-concentration molecules sensing by creation of FR angles in the suggested optical sensor. Finally, the conclusion is introduced in section 4.}
\section{Model and formalism}
\par
\emph{In our model, the Faraday effect of a 1D photonic sensing device which is schematically shown in figure 1 is investigated based on the transfer matrix method for a linearly polarized light which is composed of two decoupled circularly modes impinging normally in $z$-direction. Two graphene sheets in quantum Hall situation are covering a microfluidic channel, D, in which gas molecules and other analyzes could be absorbed to graphene's surface so that one would expect opening an electronic bandgap in its spectrum which could be sensed by measuring the corresponding FR angle. The operational principle of the device relies on the changes in the ultra-sensitive optical conductivity of graphene due to the absorption of molecules on graphene's surface. Interestingly, a short-time ultraviolet illumination could clean the graphene surface reverting it to its undoped state. Dielectric layers $Si$ and $SiO_{2}$ with dielectric constants $2.2$ and $10.3$, respectively, form an optical multilayer system around the graphene-covered microchannel as $air(AB)^{m}GDG(AB)^{m}air$ for which $m$ indicates the period of the dielectric layers around the microchannel. The FR angles are then obtained by calculating the phase difference between transmission coefficients of right- and left-handed modes. To make things as simple as possible, we consider a linear polarized light (s or p waves) normally incident on the graphene with $\varsigma_{s}=1$ and $\varsigma_{p}=-1$:
   \begin{equation}
\begin{cases}
E_{1s/p}=\varsigma_{s/p}a_{1s/p}e^{ik_{1}z}+b_{1s/p}e^{-ik_{1}z}\\
H_{1s/p}z_{1}=-a_{1s/p}e^{ik_{1}z}+\varsigma_{s/p}b_{1s/p}e^{-ik_{1}z}\\
\end{cases}
\end{equation}
Now, two circularly polarized modes in the second zone with $E_{2\pm}$ and $H_{2\pm}$ could be represented as:
\begin{equation}
\begin{cases}
E_{2\pm}=a_{2\pm}e^{ik_{2}z}+b_{2\pm}e^{-ik_{2}z}\\
H_{2\pm}=a_{2\pm}e^{ik_{2}z}+b_{2\pm}e^{-ik_{2}z}\\
\end{cases}
\end{equation}
where $k_{1} =\sqrt{n_{1}} \frac{\omega}{c}$ ($k_{2} =\sqrt{n_{2}} \frac{\omega}{c}$) is the associated wave vectors and $\omega$ and $c$ are the angular frequency and the speed of light in the vacuum, respectively. Note that $z_{i}$ (i=1,2) is assumed to stand for $z_{0}/n_{i}$  and   $a_{i\pm} = a_{ix} \pm ia_{iy}$ and $b_{i\pm} = b_{ix} \pm ib_{iy}$. From the boundary condition in the presence of a constant magnetic field which changes the optical surface conductivity of graphene one might write:
\begin{equation}
\begin{cases}
E_{2\pm}-E_{1s/p}=0\\
H_{2\pm}-H_{1s/p}=\pm iJ_{\mp}
\end{cases}
\end{equation}
 for which $J_{\pm}=\sigma_{\mp}E_{2\pm}$. To proceed, we write:
 \begin{equation}
\begin{cases}
\varsigma_{s/p}a_{1s/p}+b_{1s/p}=a_{2\pm}+b_{2\pm}\\
-a_{1s/p}+\varsigma_{s/p}b_{1s/p}=(a_{2\pm}+b_{2\pm})-\sigma_{\mp}(a_{2\pm}+b_{2\pm})\\
\end{cases}
\end{equation}
 Note that, the longitudinal parts of the optical conductivity tensor $\sigma$ for graphene which proves to be more proper for the numerical calculations are presented in [37] as:
\begin{equation}\begin{split}
\sigma_{0}(\omega)=
&\frac{e^{2}v_{f}^{2}\vert eB\vert\left(\hbar\omega+2i\Gamma \right)}{\pi i}\times\\& \sum_{n=0}^{\infty}\left\lbrace\frac{\left[f_{d}(M_{n})-f_{d}(M_{n+1})\right]+ \left[f_{d}(-M_{n+1})-f_{d}(-M_{n})\right]}{\left(M_{n+1}-M_{n}\right)^{3}-\left(\hbar\omega+2i\Gamma \right)^{2}\left(M_{n+1}-M_{n}\right)}\right\rbrace \\&
+ \left\lbrace\frac{\left[f_{d}(-M_{n})-f_{d}(M_{n+1})\right]+ \left[f_{d}(-M_{n+1})-f_{d}(M_{n})\right]}{\left(M_{n+1}+M_{n}\right)^{3}-\left(\hbar\omega+2i\Gamma \right)^{2}\left(M_{n+1}+M_{n}\right)}\right\rbrace
 \end{split}\end{equation}
 while the Hall conductivity term is:
 \begin{equation}\begin{split}
&\sigma_{H}(\omega)=\frac{-e^{2}v_{f}^{2} eB}{\pi}\\&
\sum_{n=0}^{\infty}\left\lbrace \left[f_{d}(M_{n})-f_{d}(M_{n+1})\right]- \left[f_{d}(-M_{n+1})-f_{d}(-M_{n})\right]\right\rbrace \times \\&
\left\lbrace \frac{1}{\left(M_{n+1}-M_{n}\right)^{2}-\left(\hbar\omega+2i\Gamma \right)^{2}}+ \frac{1}{\left(M_{n+1}+M_{n}\right)^{2}-\left(\hbar\omega+2i\Gamma \right)^{2}}\right\rbrace
\end{split} \end{equation}
In the above relations $M_{n} =\sqrt{\Delta^{2} +2nv_{F}^{2}|eB|h}$  shows the quantized energy levels which as it is clear could be tuned by the tunable Dirac gap. Interestingly, in our case surface transfer doping of graphene by single-molecules could correspondingly open an electronic bandgap in graphen's dispersion. Therefore, these single-molecules could be sensed by the associated measured FR angles of 1D PC. The distribution are then controlled by $f_{d}(\varepsilon)=\frac{1}{1+exp( \frac{\varepsilon-\mu_{c}}{K_{B}T)}}$ with $K_{B}$ the Boltzmann constant. The transfer matrix, $D1\rightarrow2$, yields the following relation for coupled coefficients of circularly hybrid waves as:
\begin{equation}
\begin{bmatrix}
a_{1\pm}\\
b_{1\pm}
\end{bmatrix}
=\frac{1}{2n_{1}}
\begin{bmatrix}
n_{1}+n_{2}+z_{0}\sigma_{\mp}&&
n_{1}-n_{2}+z_{0}\sigma_{\mp}\\
n_{1}-n_{2}-z_{0}\sigma_{\mp}&&
n_{1}+n_{2}-z_{0}\sigma_{\mp}
\end{bmatrix}
\begin{bmatrix}
a_{2\pm}\\
b_{2\pm}
\end{bmatrix}
\end{equation}
 Therefore, extending this relation to the introduced photonic device, one can extend it to other layers ($i - 1 \rightarrow i$; $i = 2, 3, 4, ...$) noticing that generally complex conductivity $\sigma_{±}$ can be neglected in the absence of the graphene layer. Hence, for numerical calculations we could use the following relation:
 \begin{equation}
\begin{bmatrix}
a_{1\pm}\\
b_{1\pm}
\end{bmatrix}
=\frac{1}{2n_{1}}
\begin{bmatrix}
n_{1}+n_{2}+2\alpha\sigma_{s\mp}&&
n_{1}-n_{2}+2\alpha\sigma_{s\mp}\\
n_{1}-n_{2}-2\alpha\sigma_{s\mp}&&
n_{1}+n_{2}-2\alpha\sigma_{s\mp}
\end{bmatrix}
\begin{bmatrix}
a_{2\pm}\\
b_{2\pm}
\end{bmatrix}
\end{equation}
Now, in terms of dimensionless terms which are proper for numerical calculations one can simplify more with the help of the definitions:
$\sigma_{\pm}=\sigma_{0}\sigma_{s\pm}$ and $\sigma_{0}=\frac{1}{\pi}\frac{e^2}{ h}$ to show circularly transmission coefficients for a graphene layer sandwiched between to dielectric media with refractive indices, $n_{1}$ and $n_{2}$, as:
 \begin{equation}
t_{\pm}=\frac{1}{D^{(1,1)}}=\frac{2n_{1}}{\beta+2\alpha\sigma_{s\mp}}
\end{equation}
where $\alpha=\frac{e^2}{2\pi\epsilon_{0}c h}$ is fine structure constant and $\beta=n_{1}+n_{2}$. Then, one can show that the transmission coefficients could be expressed as follows:
\begin{equation}
t_{\pm}=\frac{2n_{1}}{\gamma}[\beta+2\alpha Re(\sigma_{s\mp}) -i2\alpha Im(\sigma_{s\mp})]
\end{equation}
with $\gamma=[\beta+2\alpha Re(\sigma_{s\mp} )]^2+4\alpha^2 [Im(\sigma_{s\mp})]^2 $. Here, one can then proceed more to obtain FR angle, $\theta_{F}$, in terms of dimentionless quantities in the following expression:
\begin{equation}
2\theta_{F}=tan^{-1}[-\frac{2\alpha Im(\sigma_{s-})}{\beta+2\alpha Re(\sigma_{s-}) }]-tan^{-1}[-\frac{2\alpha Im(\sigma_{s+})}{\beta+2\alpha Re(\sigma_{s+}) }]
\end{equation}
It is clear that for calculating the FR angle of a layered structure we should also consider a propagation matrix which connects both components of light (magnetic and electric) to the neighbor dielectric layers:
\begin{equation}
P(z)=
\begin{bmatrix}
e^{-ik_{z}z}&&0\\
0&&e^{ik_{z}z}
\end{bmatrix}
\end{equation}
At this stage, for the layers structure acting in infrared zone, gathering of both transmission matrix and propagation one for the first and the last layers gives :
\begin{equation}
H=M^{(1,2)}P(x^{(1,2)})M^{(2,3)}P(x^{(2,3)})....P(x^{(N-1,N)})M^{(N,N+1)}
\end{equation}
The transmission probability for both circularly polarized (left- and right handed) light are calculated as:
\begin{equation}
T_{\pm}=\frac{1}{H^{(1,1)}}= \vert t_{\pm}\vert^{2}
\end{equation}
Finally, the transmission probability of the transmitted light is obtained as:
\begin{equation}
T=\frac{n_{2}}{2n_{1}}(T_{+}+T_{-})
\end{equation}
from which FR angle is also calculated based on $\theta_{\pm}$, i.e. the phase of the left- and right-handed modes as:
\begin{equation}
\theta_{F}=\frac{1}{2}(\theta_{+}-\theta_{-})
\end{equation}
}
\section{Numerical results}
\par
\emph{Considering the central wavelength $\lambda_{0}=143\ \mu m$, in this section, we numerically calculate FR angles and also the transmission probability of circular microcavity defect modes at room temperature for low values of the Fermi energy ranging mostly from $\mu=0.1\ meV$ to $\mu=0.9\ meV$ which is almost $1000$ times smaller than previous works studying the optical properties of graphene-based sensors. These low Fermi energy changes allow one to detect low-concentrations of molecules attached to the graphene's surface. Moreover, the sign of these changes in the chemical potential could help to indicate the p-type or n-type doping of attached molecules. We first, as our main aim in this work, examine the effect of a gap opening on the FR angles of the linearly polarized defect modes of the suggested optical sensing structure. As it is revealed in Fig. 2 for $B=10 \ T$ and $\mu=0.9 \ meV$, opening a Dirac gap in graphene's spectrum could significantly affect FR angles. In fact, increasing the electronic gap at Dirac points in graphene's spectrum by attaching low-weight molecules to its surfaces shows to decrease the rotation of the light's polarization so that, as it is guessed from the plots, for $\Delta=200\ meV$ and beyond almost no FR effect is expected to be observed. \\
In Fig. 3, the effect of both positive and negative chemical potential on the FR angles of the defective photonic structure is presented. It is clear that for a constant magnetic field, $B=10 \ T$, and $\Delta=0$ increasing the chemical potentials gives stronger rotations to the polarization's plane of the defect states. Note that, as it is observed, negative values of the chemical potential cause the linearly polarized light to rotate in an opposite direction which means that for p-type transfer doping one gets opposite signs for the rotation angles of the linearly polarized defect modes. Interestingly, however, we see that the transmission of the defect modes increases as Dirac gaps open wider and wider. Therefore, as a matter of more illustration, this behavior is depicted in a 3D plot (see Fig. 4) for opening gaps ranging from $1$ to $150\ meV$. Interestingly, we see that in the proposed interval for the electronic bandgap, the transmission probability remains relatively constant. However, the FR angles exhibit significant changes by increasing the gap values.  \\
In the next step, in Fig. 5, we simulate the behavior of resonance defect modes for a constant opening Dirac gap of $50\ meV$ and applied magnetic fields $B=5$, $B=10$ and $B=20\ T$. Surprisingly, it is seen that stronger fields could decrease the rotations of the polarization plane of light in some cases. However, this behavior is not generally true for the transmission of microcavity resonance modes as it is illustrated. Generally, however, we see that the effect of increasing the magnetic field on the FR angles is shifting the angle of rotations toward the central wavelength $\lambda_{0}=143\ \mu m$. The reason behind shifting the rotation angle toward central frequency modes is that increasing the intensity of the magnetic field causes separated left- and right-handed circular modes of the linearly light to move toward merging in the frequency space.
\\
Furthermore, in Fig. 6, we consider the FR effect for different values of the chemical potential ranging from $0.1$ to $0.9\ meV$ in the case of gap openings $\Delta=5$ and $\Delta=50\ meV$. Here, we could see that the modulation of the chemical potential toward higher values leads to stronger rotations of the polarization's plane of the light at a constant magnetic field $B=10\ T$. It is also obvious that in the case of wider opening for Dirac gaps in graphene's spectrum the rotations are observed to be shifted slightly toward the central wavelength. This situation is better visible fin a 3D view in Fig. 7 for which it is clear that the rotation of the polarization's plane of resonance modes shows a linear change relative to the changes in the value of the chemical potential. It also clear that for higher values of the chemical potential the rotations of the resonance modes are more affected by the gap opening.
\\
At this point, we also turn our attention to the case of adding some layers to our structure. Fig. 8 depicts the situation for the effect of increasing the gap value on the FR angles in the case of m=7 and a constant magnetic field $B=10\ T$ at the room temperature. We see that compared to the situations shown in Fig. 2 we now have stronger rotations. One also could notice the role of the opening gap in merging the two left- and right-handed modes.
\\
 At the end of this section, as a matter of more comparison, we consider the situation when one of the graphene layers is removed. Interestingly, as is illustrated in figure 9, both transmission probability and FR angle shown to be decreased. Particularly, for an opened electronic bandgap in graphene's spectrum, $\Delta=60\ meV$ we see that the removal of one of the graphene layers leads to lower FR angles. On the other hand, as it might have been expected one gets stronger transmission for the single covering the case of the microchannel by graphene.}
\section{Conclusion}
 \par
\emph{To investigate the sensing properties of the Faraday effect in graphene, we considered the modulations of the energy bandgap of graphene when gas molecules are adsorbed to its surface. In this work, we numerically investigated the effect of a tunable electronic gap opening in graphene's spectrum on FR rotation for sensing applications of a 1D graphene-based defective PC. These induced electronic bandgaps could be opened in graphene's linear dispersion while single gas molecules are attached to its surface [38]. We demonstrated that the presence of these Dirac gaps affect significantly FR angles corresponding to the light prorogation into the suggested photonic sensing device due to the nonlinear effects induced in the energy spectrum of graphene at Dirac points. On-line monitoring was then proved to be possible by considering the sign and amount of FR angles in a constant magnetic field applied on gapped graphene layers for low chemical potentials. To be more specific, in this work, we first obtained the transmission and the Faraday effect for the proposed 1D structure with a microfluidic channel covered by gapped graphene layers in the case of relatively very low changes in the value of the Fermi energy which could be due to the single molecular absorption on the graphene's surface. The existence of an opening gap at Dirac points in graphene's spectrum was shown to decrease significantly FR rotations while keeping the transmission almost unchanged in the proposed photonic sensing structure. Also, in addition to advances in the electronic bandgap engineering of graphene, a gap in graphene's spectrum could be created by applying strong magnetic fields( excitonic gap) [39]. Therefore, FR angles studied in this work could provide an optical contact-free method to measure and sense corresponding opening bandgaps created in the linear graphene's energy dispersion relation under different situations. Moreover, it should be pointed out that all the numerical calculations were carried out at room temperature and relatively low magnetic fields which could open routes towards the use of the introduced sensing layered device in real-world applications.\\
   Significantly, we showed that the FR angle for a linearly polarized light shows narrower bandwidth upon increasing the period of the 1D photonic structure and also gets relatively higher values by increasing the opening bandgap. To be more specific, it was observed that inducing an electronic bandgap $\Delta=50 meV$ in graphene's spectrum for the period $m=7$ gives stronger FR angles relative to the case for which $m=5$. More importantly, preservations of Dirac cones upon chemical doping in a very short time increase the chemical potential of graphene i.e. $\mu=E_{F}-E_{D}$ and thus an FR occurs for low weight-molecules absorption on graphene in a very short time could be detected. Therefore, relatively short response times in sensing applications are expected for the proposed 1D defective photonic structure. In the end, we showed that the existence of both graphene layers covering the microchannel cavity is essential since, as we showed numerically, the removal of one of the graphene layers decreased the FR angles significantly.
\\
  In summary, the existence of the bandgap-tunable Faraday effect in a 1D graphene-based defective optical sensing structure was numerically demonstrated. Moreover, sensing properties of the introduced model upon different surface transfer doping - which could lead to opening a Dirac gap in graphene's linear spectrum - was studied.}

%\end{multicols}

 \begin{figure}
\begin{center}
\includegraphics[width=18cm]{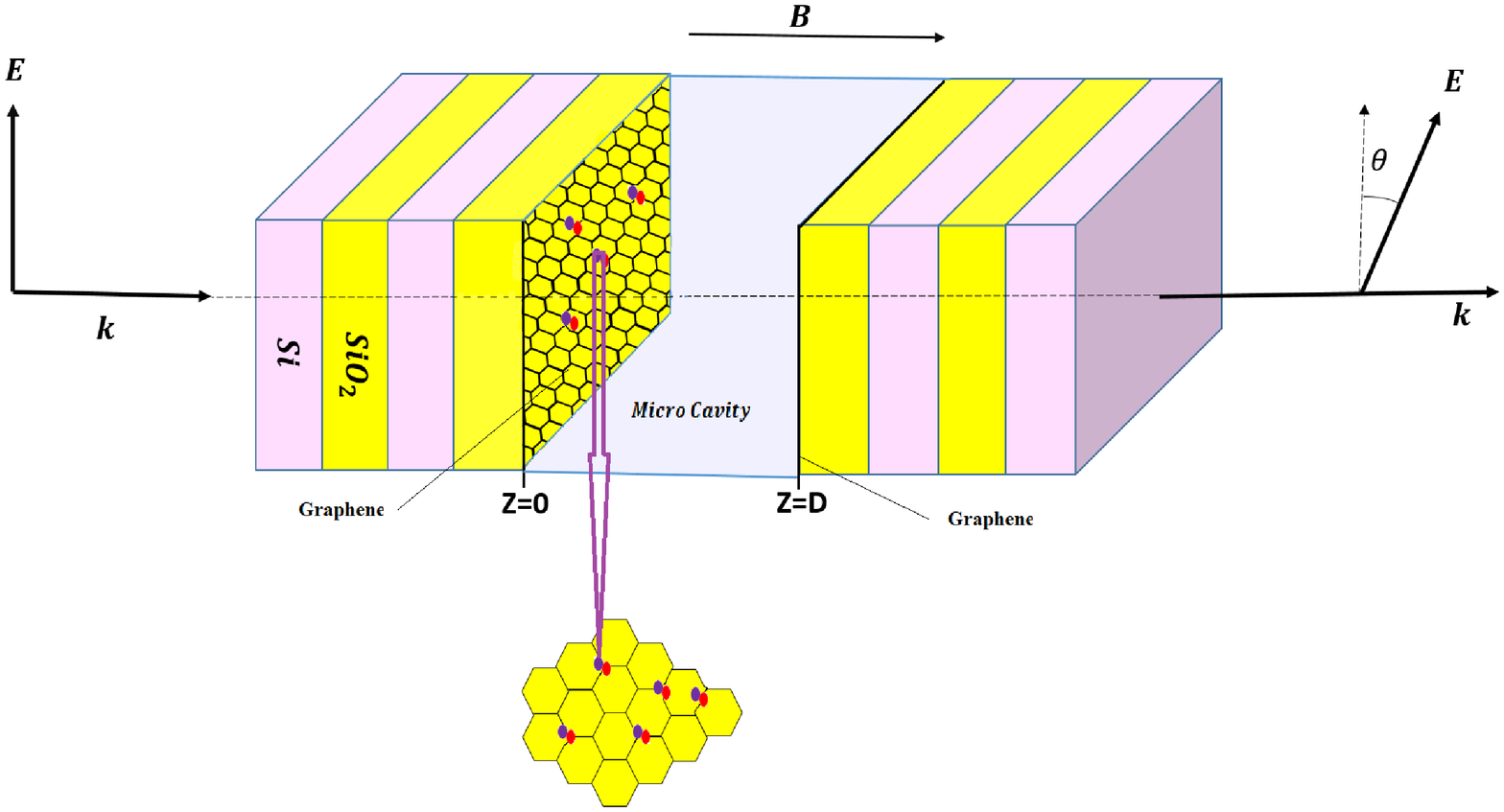}
\caption{ The proposed defective optical sensing device with a fluidic microchannel covered by graphene layers. The changes in FR angles could be observed by attaching low weight molecules to gapped-graphene's surface.}
\end{center}
\end{figure}
\begin{figure}
\begin{center}
\includegraphics[width=18cm]{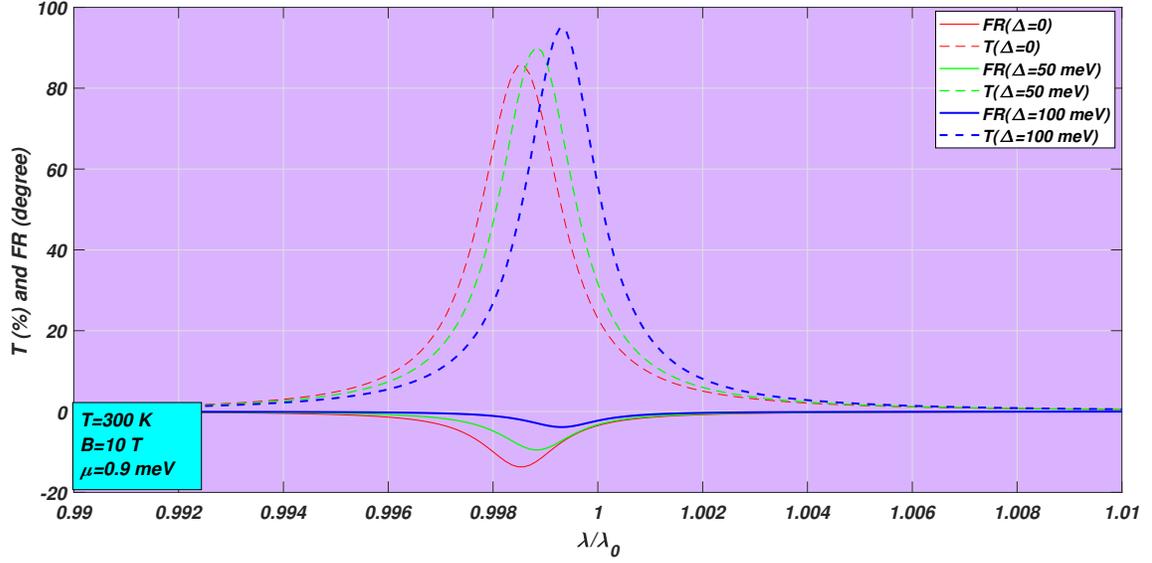}
\caption{Effect of gap opening at Dirac points in graphene on Faraday modes and the corresponding transmission of EM waves. It is obvious that there is a trade-off between the values of electronic gaps and FR effect in the introduced optical sensing structure with the period number $m=5$.  }
\end{center}
\end{figure}
\begin{figure}
\begin{center}
\includegraphics[width=18cm]{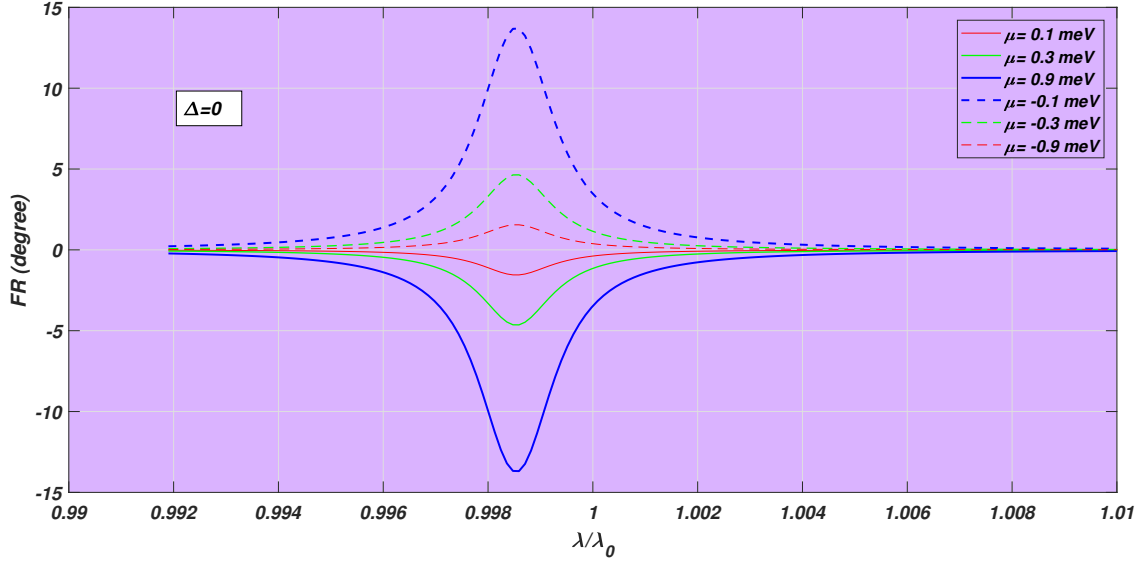}
\caption{Faraday modes for gapless graphene are shown to posses both positive and negative signs which corresponds to the two direction of FR angles for the different doping transfer on the graphene's surface due to changes in the sign of the chemical potential. It also observed that increasing the chemical potential leads to the observation of stronger rotations of the polarization's plane of EM modes.}
\end{center}
\end{figure}
\begin{figure}
\begin{center}
\includegraphics[width=18cm]{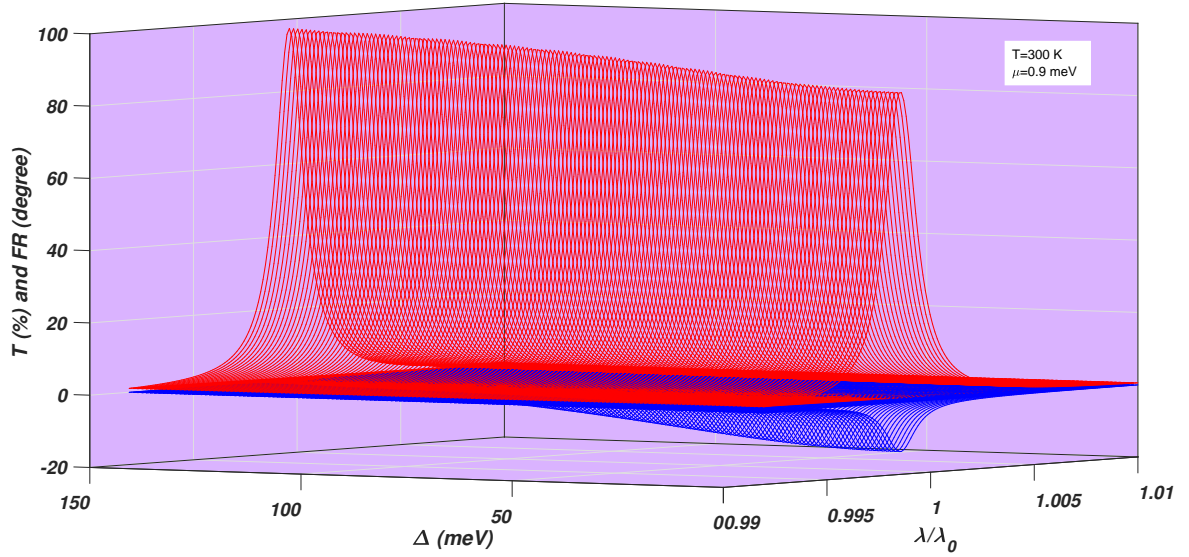}
\caption{Effect of gap opening $\Delta$ at Driac points in graphene's energy dispersion on FR angles and also the transmission modes for a constant chemical potential $\mu=0.9\ meV$. Interestingly, as it is seen, in the interval $\Delta=0$ to $\Delta=150\ meV$ for the electronic bandgap, FR angles show to be more sensitive to the changes of $\Delta$ than the transmission probability which is relatively constant.}
\end{center}
\end{figure}
\begin{figure}
\begin{center}
\includegraphics[width=18cm]{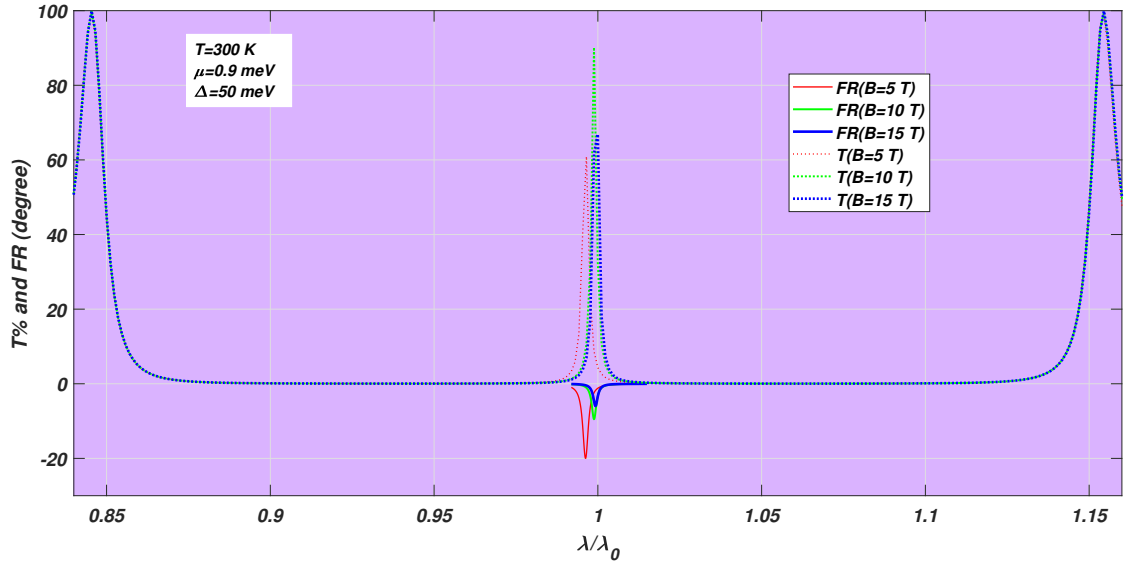}
\caption{The frequency shift of FR angles for two different Dirac gaps is illustrated. It is also shown that the effect of increasing the chemical potential for the both situation leads to no frequency shift for FR angles.}
\end{center}
\end{figure}
\begin{figure}
\begin{center}
\includegraphics[width=18cm]{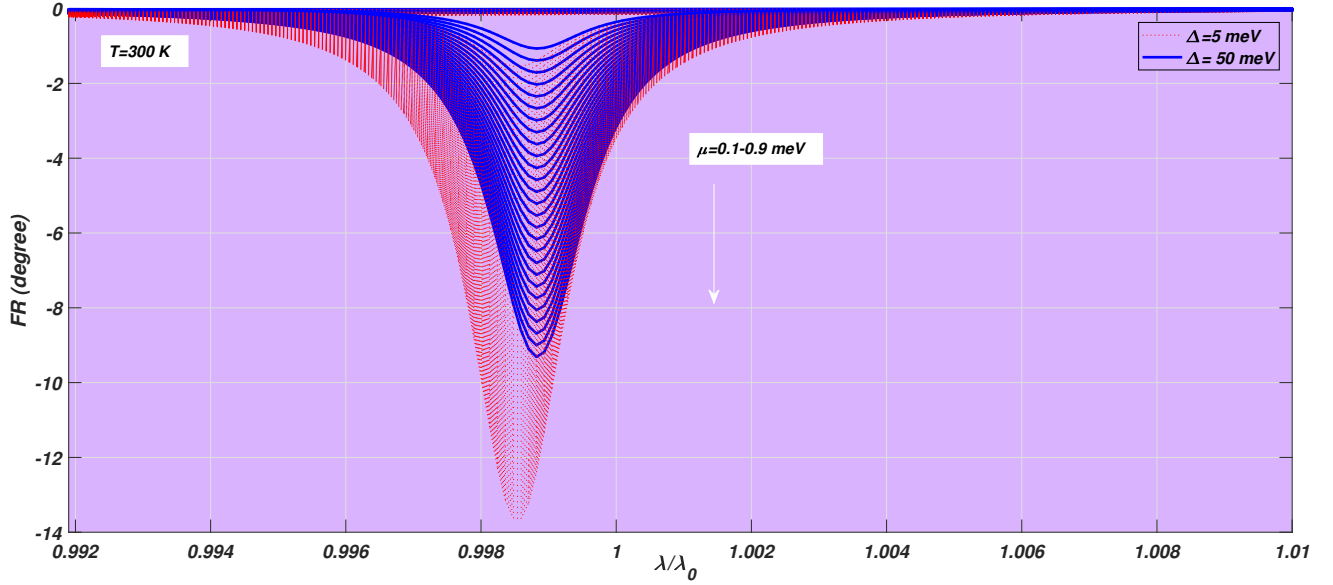}
\caption{The effect of increasing chemical potential at $B=10\ T$ for gaps $\Delta=5$ and $\Delta=50\ meV$. The blue shift of the modes are observed for the higher Dirac opening.}
\end{center}
\end{figure}
\begin{figure}
\begin{center}
\includegraphics[width=18cm]{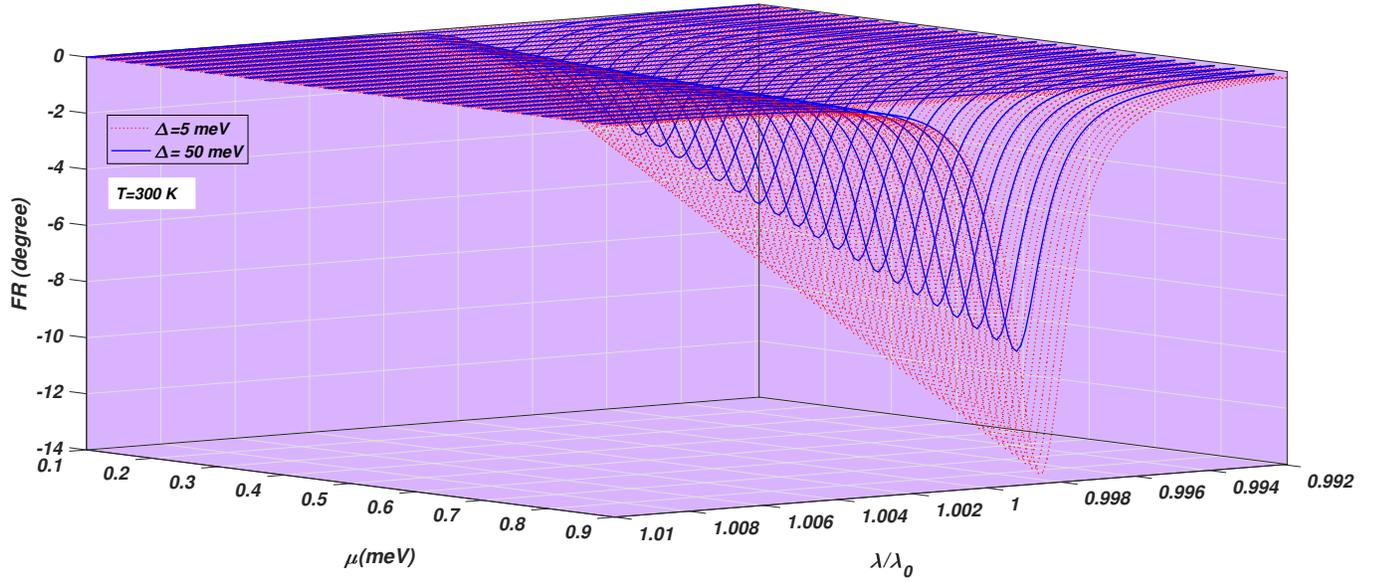}
\caption{Faraday effect for different values of Dirac gap opening at room temperature. Linear relation between FR angles and the value of the chemical potential is observed in this 3D plot for two different $\Delta=5\ meV$ and $\Delta=50\ meV$. Significantly, it is seen that the separation of FR modes for the two values occurs at higher chemical potentials. }
\end{center}
\end{figure}
\begin{figure}
\begin{center}
\includegraphics[width=18cm]{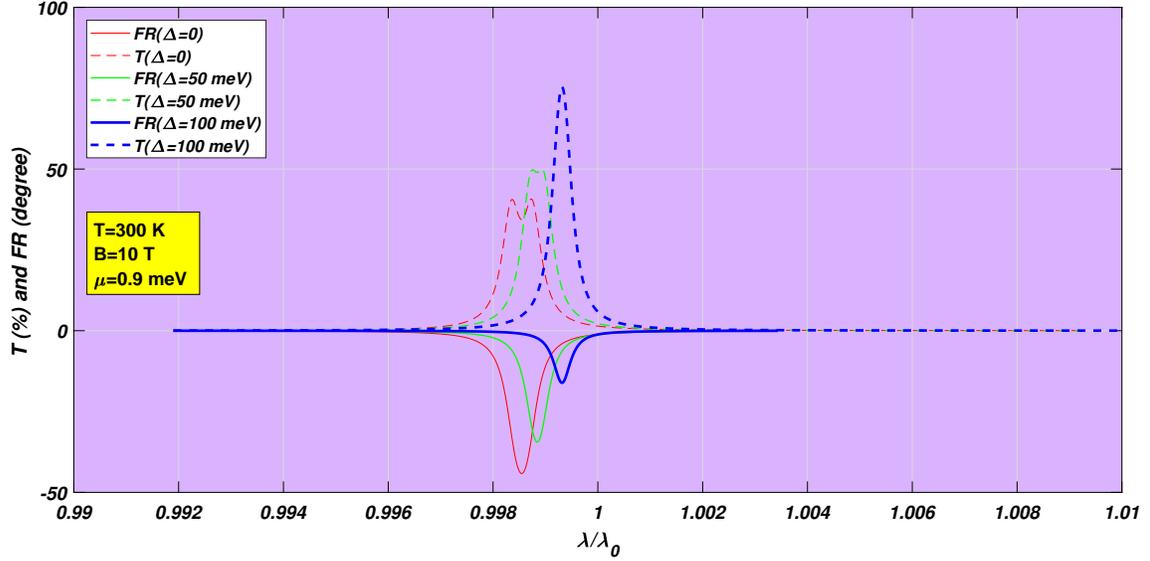}
\caption{Effect of gap opening for Fr angles and their corresponding transmissions at the room temperature for an applied magnetic field $B=10\ T$. As it seen Faraday effect is more diminished upon higher bandgaps in graphene's spectrum.}
\end{center}
\end{figure}
\begin{figure}
\begin{center}
\includegraphics[width=18cm]{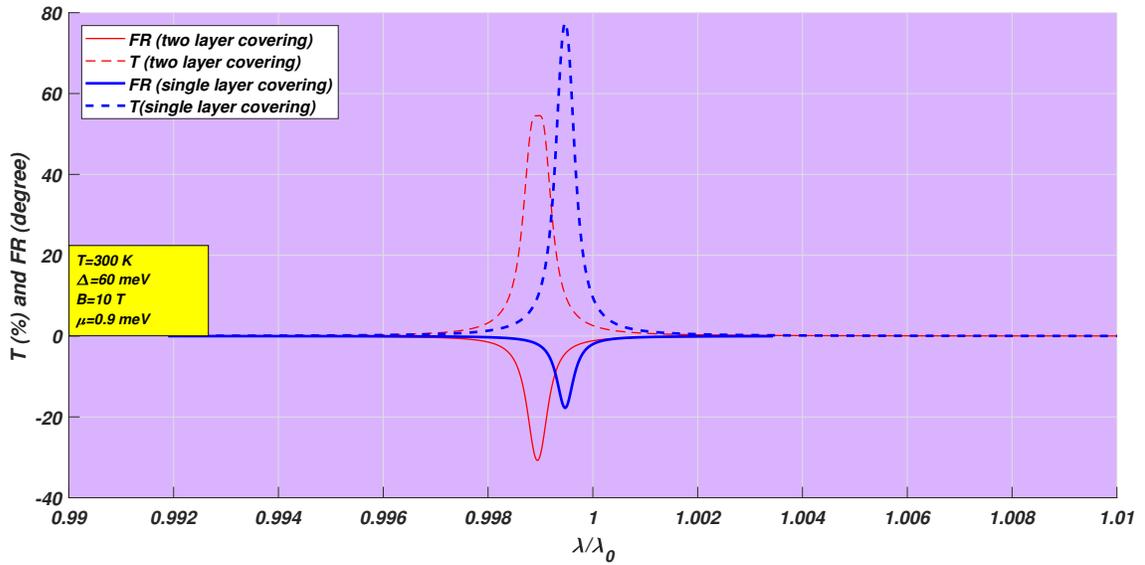}
\caption{For an opened electronic bandgap, $\Delta=60\ meV$ we see that the removal of one graphene layer leads to lower FR angles. On the other hand, as it might have been expected one gets stronger transmission for the single covering case of the microchannel by graphene.}
\end{center}
\end{figure}

\end{document}